\documentclass[a4paper,twocolumn,english,showpacs,aps,prl,fleqn,superscriptaddress]{revtex4-1}

\usepackage[T1]{fontenc}
\usepackage[utf8]{luainputenc}
\usepackage{amsmath}
\usepackage{amssymb}
\usepackage{graphicx}
\usepackage{hyperref}
\usepackage{babel}
\usepackage{wasysym}
\usepackage{microtype}
\usepackage{color}
\usepackage{textcomp}

\begin{document}
\renewcommand\figurename{FIG.}
\title{Observation of state-to-state hyperfine-changing collision in a Bose-Fermi mixture of $^6$Li and $^{41}$K atoms}

\author{Xiao-Qiong Wang}
\affiliation{Shanghai Branch, National Laboratory for Physical Sciences at Microscale and Department of Modern Physics, University of Science and Technology of China, Shanghai, 201315, China}
\affiliation{CAS Center for Excellence and Synergetic Innovation Center in Quantum Information and Quantum Physics, University of Science and Technology of China, Hefei, Anhui 230026, China}

\author{Yu-Xuan Wang}
\affiliation{Shanghai Branch, National Laboratory for Physical Sciences at Microscale and Department of Modern Physics, University of Science and Technology of China, Shanghai, 201315, China}
\affiliation{CAS Center for Excellence and Synergetic Innovation Center in Quantum Information and Quantum Physics, University of Science and Technology of China, Hefei, Anhui 230026, China}

\author{Xiang-Pei Liu}
\affiliation{Shanghai Branch, National Laboratory for Physical Sciences at Microscale and Department of Modern Physics, University of Science and Technology of China, Shanghai, 201315, China}
\affiliation{CAS Center for Excellence and Synergetic Innovation Center in Quantum Information and Quantum Physics, University of Science and Technology of China, Hefei, Anhui 230026, China}

\author{Ran Qi}
\affiliation{Department of Physics, Renmin University of China, Beijing, 100872, China}

\author{Xing-Can Yao}
\affiliation{Shanghai Branch, National Laboratory for Physical Sciences at Microscale and Department of Modern Physics, University of Science and Technology of China, Shanghai, 201315, China}
\affiliation{CAS Center for Excellence and Synergetic Innovation Center in Quantum Information and Quantum Physics, University of Science and Technology of China, Hefei, Anhui 230026, China}

\author{Yu-Ao Chen}
\affiliation{Shanghai Branch, National Laboratory for Physical Sciences at Microscale and Department of Modern Physics, University of Science and Technology of China, Shanghai, 201315, China}
\affiliation{CAS Center for Excellence and Synergetic Innovation Center in Quantum Information and Quantum Physics, University of Science and Technology of China, Hefei, Anhui 230026, China}

\author{Jian-Wei Pan}
\affiliation{Shanghai Branch, National Laboratory for Physical Sciences at Microscale and Department of Modern Physics, University of Science and Technology of China, Shanghai, 201315, China}
\affiliation{CAS Center for Excellence and Synergetic Innovation Center in Quantum Information and Quantum Physics, University of Science and Technology of China, Hefei, Anhui 230026, China}


\begin{abstract}
Hyperfine-changing collisions are of fundamental interest for the studying of ultracold heteronuclear mixtures. Here, we report the state-to-state study of the hyperfine-changing-collision dynamics in a Bose-Fermi mixture of $^6$Li and $^{41}$K atoms. The collision products are directly observed and the spin-changing dynamics is measured. Based on a two-body collision model, the experimental results are simultaneously fitted from which the spin-changing rate coefficient of $ 1.9(2)\times 10^{-12}~\rm{cm^3\cdot s^{-1}}$ is gained, being consistent with the multi-channel quantum defect theory calculation. We further show that the contact parameter of $^6$Li-$^{41}$K mixture can be extracted from the measured spin-changing dynamics. The obtained results are consistent with the first order perturbation theory in the weakly-interacting limit. Our system offers great promise for studying spin-changing interactions in heteronuclear mixtures.

\end{abstract}

\maketitle

\date{\today}

Ultracold Bose-Fermi mixtures provide an ideal platform for studying few-body~\cite{Chin2010} and many-body physics~\cite{Bloch2008}, particularly due to the distinct properties of the two atomic species. Over the past decades, many important quantum phenomena arising from the Bose-Fermi interactions have been demonstrated, such as Fermionic polar molecules~\cite{Ni2008, Wu2012}, heteronuclear Efimov states~\cite{Tung2014, Pires2014}, Bose polarons~\cite{Hu2016}, and Bose-Fermi double superfluidity~\cite{Ferrier-Barbut2014,Yao2016,Wu2018}. Moreover, Bose-Fermi mixtures are natural systems to realize the Kondo effect~\cite{Bauer2013}, since the exchange interaction can be mediated by spin-changing collisions between the two species. However, due to their complex scattering potentials, inelastic spin-changing collisions are ubiquitous in a Bose-Fermi mixtures, resulting in unavoidable heating, decoherence, and reduced lifetime. Therefore, understanding the inelastic spin-changing collisions between two distinct atomic species is crucial not only for the realization of quantum degenerate mixtures, but also for the study of novel quantum phases therein.

To date, hyperfine-changing inelastic collisions (HCC) have been studied in various heteronuclear mixture systems, such as Li-Cs~\cite{Mudrich2004}, Na-K~\cite{Santos1999}, Na-Rb~\cite{Young2000}, Rb-K~\cite{Marcassa2000} and mixture of Rubidium isotopes~\cite{Xu2015}. In these studies, the products of HCC are hardly observed, since the released energy is much larger than the trap depth. Therefore, the HCC have been usually studied through the measurement of the overall loss rates of the atoms in the initial hyperfine manifold. While challenging, the ability to directly measure the evolution of initial and final spin states will provide a powerful tool for studying HCC, allowing the comparison between experimental results and theoretical predictions for deterministic scattering channels. Furthermore, the state-to-state HCC offers great opportunities to obtain the contact parameter, which is a central quantity that connects the two-body short range correlations to the macroscopic thermodynamic quantities through a set of universal relations~\cite{Tan2008a,Tan2008,Tan2008b,Braaten2008,Braaten2010}. While great efforts have been devoted to probing the contact for both the Bose~\cite{Wild2012,Fletcher2017} and Fermi gases~\cite{Stewart2010,Sagi2012,Carcy2019}, few of them are on the Bose-Fermi mixtures.

\begin{figure}
  \centering
  \includegraphics[width=\columnwidth]{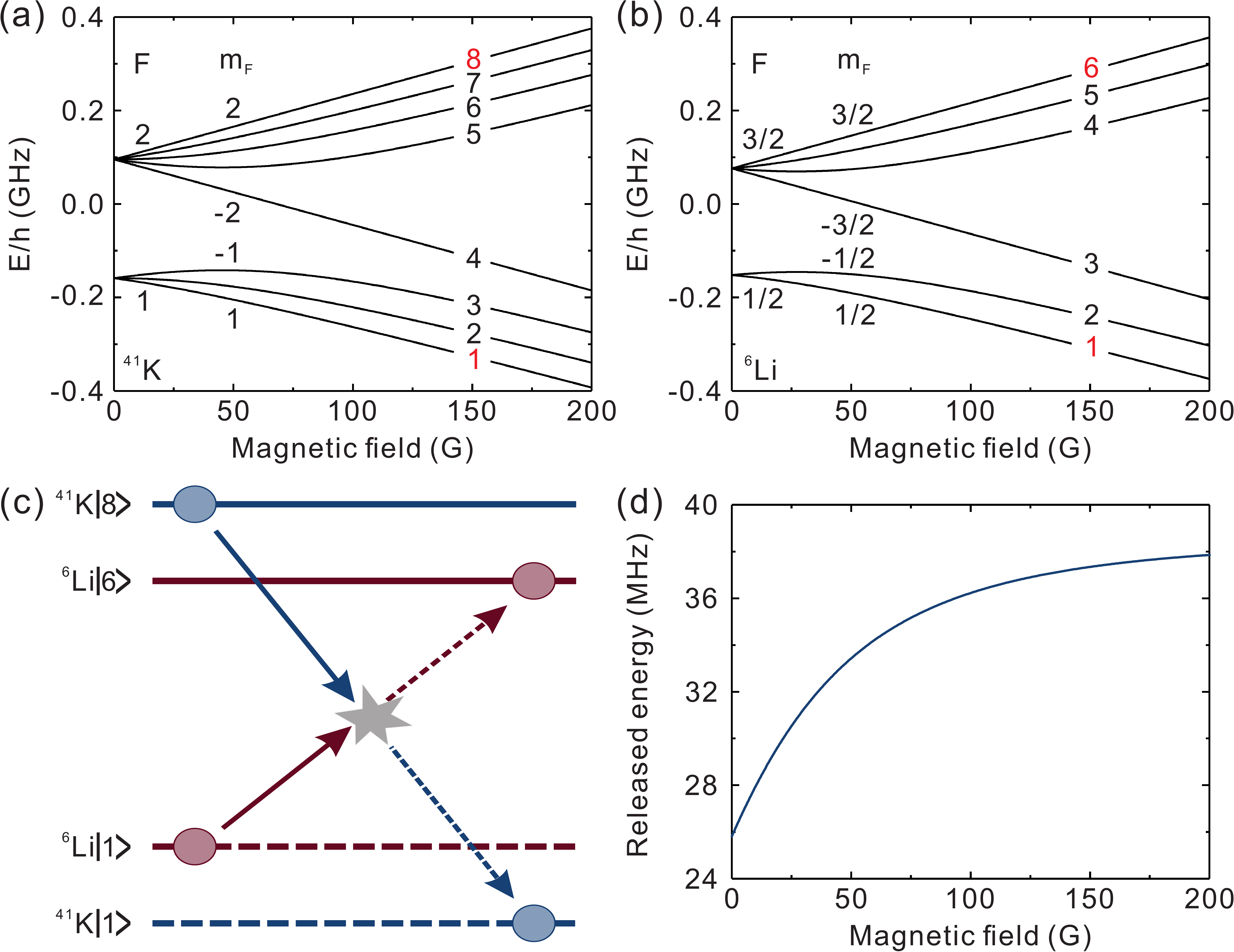}
  \caption{(a) and (b) are the magnetic-field dependence of ground states energies of $^{41}$K and $^6$Li atoms, respectively. The zeeman levels are labeled with numbers and the states of interest are marked as red. (c) is the sketch of the HCC between $^{41}$K and $^6$Li atoms. Blue (red) solid and dashed lines represent the highest and lowest Zeeman states of $^{41}$K ($^6$Li) atoms, respectively. Blue and red dots correspond to $^{41}$K and $^6$Li atoms, respectively. (d) is the magnetic-field dependence of energy difference between the initial and final channels of the HCC.}\label{figure1}
\end{figure}

In this Letter, we report the observation of state-to-state HCC dynamics of an ultracold $^6$Li-$^{41}$K Bose-Fermi mixture confined in a cigar-shaped optical dipole trap.  The HCC occurs between $^6$Li and $^{41}$K atoms in the lowest and highest zeeman ground states, respectively. Due to the conservation of total magnetic quantum number $m_F$, this process can be written as $|1\rangle_{Li}+|8\rangle_{K} \rightarrow |6\rangle_{Li}+|1\rangle_{K}$, where $|i\rangle$ represents $|F, m_F\rangle$ state (see Fig.~\ref{figure1}(a)-(b) for the interested zeeman states). By directly probing the $^6$Li and $^{41}$K atoms in different zeeman states, we measure the increase and loss of $^{41}$K atoms in the final and initial states, respectively. The loss of $^6$Li atoms in the initial state is probed accordingly. The spin-changing coefficient $L_2$ of $ 1.9(2)\times 10^{-12}~\rm{cm^3\cdot s^{-1}}$ is obtained by simultaneously fitting the experimental data. A multi-channel quantum defect theory (MQDT) calculation \cite{Gao1998, Gao2005, Gao2008} is performed, and reasonable agreement between experiment and theory is achieved. Furthermore, by using the calculated imaginary part of the $^6$Li-$^{41}$K scattering length, the contact parameter of $|1\rangle_{Li}$-$|8\rangle_K$ ($C_{|1\rangle_{Li}-|8\rangle_K}$) is successfully extracted from the loss dynamics of $|8\rangle_{K}$.

The experimental procedure for preparing the $^6$Li-$^{41}$K mixture is similar to our previous works~\cite{Chen2016, Yao2016, Wu2017}. After the laser cooling and magnetic transport phase, both the $^6$Li and $^{41}$K atoms are confined in an optically-plugged magnetic trap for evaporative cooling. Then, we transfer the mixture into a cigar-shaped optical dipole trap (wavelength 1064~nm, $1/e^2$ radius 32(1)$~\mu$m), where $1.27(4)\times 10^7$ $^6$Li atoms in the state $|6\rangle_{Li}$ and $6.5(2)\times 10^6$ $^{41}$K atoms in the state $|8\rangle_{K}$ are obtained, respectively. The radial and axial trapping frequencies of $^6$Li  atoms are 2$\pi \times$6.4(4)~kHz and 2$\pi \times$48(5)~Hz, while those of  $^{41}$K atoms are 2$\pi \times$3.6(2)~kHz and  2$\pi \times$27(3)~Hz, respectively.

To investigate the HCC dynamics, the $^6$Li atoms should be prepared to $|1\rangle_{Li}$ while the $^{41}$K atoms remain in $|8\rangle_{K}$. In the experiment, a 10~ms Landau-Zener sweep ($LZ_{Li6\rightarrow Li1}$) at 46.47(1)~G is applied to transfer $^6$Li atoms from $|6\rangle_{Li}$ to $|1\rangle_{Li}$. Strikingly, about 30\% $^{41}$K atoms are flipped from $|8\rangle_K$ to $|1\rangle_K$ simultaneously. We mention that the frequency of $LZ_{Li6\rightarrow Li1}$ is far off-resonant from that of $LZ_{K8\rightarrow K1}$ and thus cannot be responsible for this spin changing. This is further verified by applying a same $LZ_{Li6\rightarrow Li1}$ only for the $^{41}$K atoms, and no such spin-changing is observed. Therefore, we attribute the spin-changing of $^{41}$K atoms to the $|1\rangle_{Li}+|8\rangle_{K} \rightarrow |6\rangle_{Li}+|1\rangle_{K}$ inelastic collisions (see Fig.~\ref{figure1}(c) for illustration).

\begin{figure}
  \centering
  \includegraphics[width=\columnwidth]{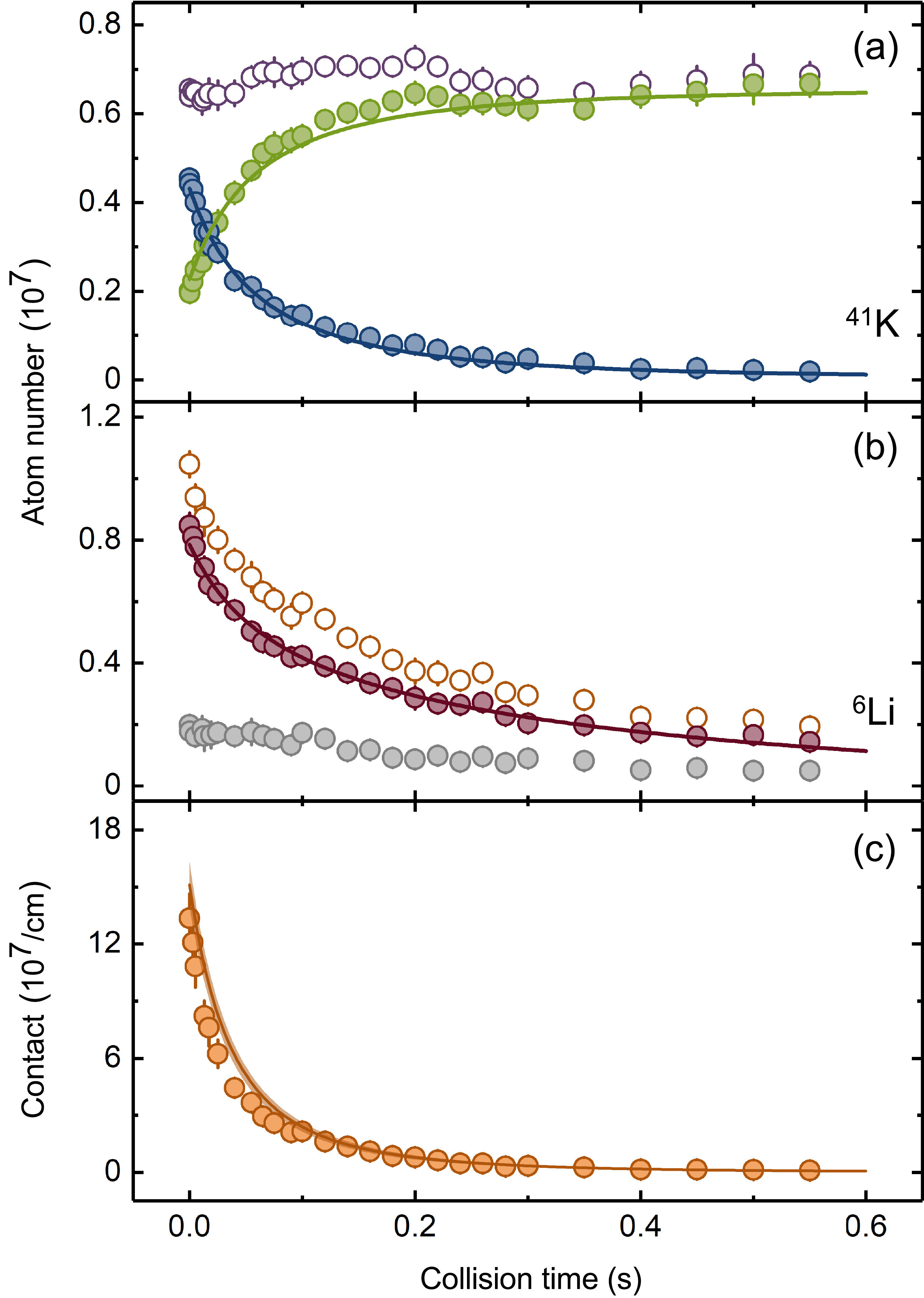}
  \caption{The HCC dynamics of $^6$Li-$^{41}$K mixture in an optical dipole trap. (a) $^{41}$K atom number as a function of collision time. Blue and green solid dots represent atom number in state $|8\rangle_K$ and $|1\rangle_K$, respectively. Purple circles corresponds to the total $^{41}$K atom number. (b) $^6$Li atom number as a function of collision time. Gray and red dots are atom number in state $|6\rangle_{Li}$ and $|1\rangle_{Li}$, respectively. Orange circles represent the total atom number of $^6$Li. Solid lines are the fitting curves based on two-body loss equations (4)-(6). All the data points and error bars are the average and standard deviation of four measurements. (c) The contact parameter of $|1\rangle_{Li}-|8\rangle_K$ mixture as a function of collision time. Orange line with shaded error region is numerically calculated from the fitting curve of the measured spin-changing dynamics, based on Eq.~\ref{eq7}. Orange dots correspond to contact parameter at weakly-interacting limit, which is calculated from Eq.~\ref{eq10}.}\label{figure2}
\end{figure}

Next, we aim to directly probe the HCC by measuring the collision products. Typically, the HCC is an exoergic process during which the released energy converts to thermal energy of the cloud. This results in heating induced atom loss and prevents the observation of collision products. Fortunately, $^6$Li-$^{41}$K mixture possesses several advantages that allow the observation of state-to-state HCC. First, the released energy, which corresponds to the energy difference between the initial and final states of the HCC, is much smaller than that of previously-studied mixtures. For instance, at a magnetic field of 46.47~G, the released energy is about 33~MHz or 1.58~mK (see Fig.~\ref{figure1} (d)). Second, due to the large mass imbalance between $^6$Li and $^{41}$K atoms, most of the energy (1.38~mK) is absorbed by $^6$Li atoms, while only a small percentage of that (200~$\rm{\mu K}$) is absorbed by $^{41}$K atoms. Third, the trapping potential for the two species is also imbalanced, which are $U_{0}^{Li}=298(20)~\rm{\mu K}$ and $U_{0}^K=652(43)~\rm{\mu K}$, respectively. Consequently, although the resultant $^6$Li atoms are ejected from the trap, the collided $^{41}$K atoms are still confined and thus can be observed.

\begin{figure}
  \centering
  \includegraphics[width=\columnwidth]{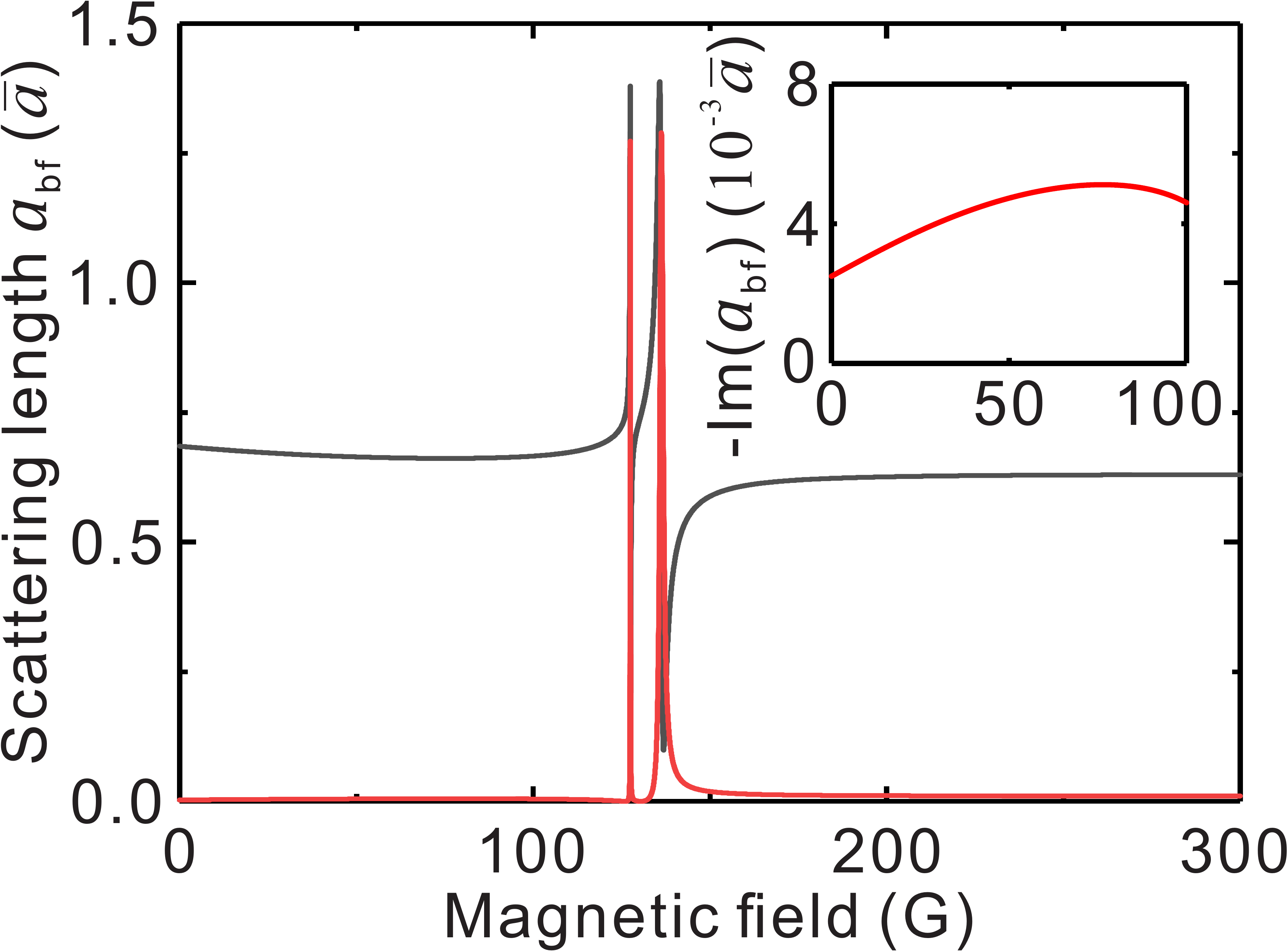}
  \caption{Scattering length of $|8\rangle_K$-$|1\rangle_{Li}$, obtained from MQDT calculation. Black and red lines correspond to Re($a_{bf}$) and $-$Im($a_{bf}$), respectively. $\overline{a}=0.956R_{vdw}$, where $R_{vdw}=40.8a_0$ is the $^6$Li-$^{41}$K van der Waals length. The inset shows $-$Im($a_{bf}$) in the range of 0 $\sim$ 100 G.}\label{figure3}
\end{figure}

In the experiment, the atom numbers of $|1\rangle_K$ and $|1\rangle_{Li}$ are simultaneously detected after a variable collision time. Furthermore, the same measurements for $|8\rangle_K$ and $|6\rangle_{Li}$ are also implemented for complementary purpose. To suppress the systematic errors, these two measurements are performed alternately. We mention that the imaging of atoms is crucial in these measurements. Typically, due to the almost closed imaging transitions in the Paschen-Back regime, the atom numbers in different Zeeman levels can be accurately obtained by using high-field imaging. Unfortunately, in our system, high-field imaging is not suitable for studying the dynamics due to the following two reasons. First, 20~ms is required for ramping the magnetic field to about 400~G, which is not favorable for studying the dynamics.  Second, a Feshbach resonance between $|8\rangle_K$ and $|1\rangle_{Li}$ atoms exists at a magnetic field of about 127~G (see Fig.~\ref{figure3}), which makes the data analysis complicated. To solve these problems, we first probe the cloud at high magnetic field to confirm that only the four interested spin states (see Fig.~\ref{figure1}(c)) are occupied during the HCC. Then, the cloud are detected at 46.47~G and 400~G with resonant optical transition of each spin state, respectively, to calibrate the imaging parameters at 46.47~G. With these two steps, we can directly probe the lowest and highest energy states of the two species at 46.47~G and obtain their corresponding atom numbers.

The experimental results are shown in Fig.~\ref{figure2} (a)-(b), where the atom numbers of four interested spin states are plotted as a function of collision time. It's seen that the $|8\rangle_K$ and $|1\rangle_{Li}$ atoms gradually decrease with collision time. On the contrary, the $|1\rangle_K$ atoms gradually increase with collision time. In addition, the decay of total  $^{41}$K atoms during the HCC is hardly visible, while about 80\% $^6$Li atoms are ejected from the trap due to heating. These behaviours qualitatively reflect the dynamics of spin-changing collision. To extract quantitatively information from the measurements, such as the spin-changing rate coefficient $L_2$, we attempt to apply a simple two-body collision model to describe these curves. We mention that the duration of spin-changing dynamics is far less than the system lifetime and the background scattering length is very small, thus the one-body and three-body process can be safely neglected.

Considering a two-body process, the local density $n(\textbf{r})$ of an atom cloud has the following rate equation:

\begin{eqnarray}
\frac{dn(\textbf{r})}{dt}=-L_2n^2(\textbf{r}).\label{eq1}
\end{eqnarray}
Then, we can obtain the rate equation of the atom number by integrating the above equation over the whole space as follow:

\begin{eqnarray}
\frac{dN}{dt}=-L_2\int n^2(\textbf{r})d^3r.\label{eq2}
\end{eqnarray}

Since the $^6$Li-$^{41}$K mixtures are thermal clouds, their density distributions n(\textbf{r}) obey the Boltzmann distribution:
\begin{eqnarray}
\begin{split}
n_{\sigma}(\textbf{r})=\frac{N_\sigma}{\pi^{3/2}\prod_{i=x,y,z} b_{i\sigma}}exp(-\sum_{i=x,y,z} \frac{i^2}{b_{i\sigma}^2}),\label{eq6}
\end{split}
\end{eqnarray}
where $b_{i\sigma} =\sqrt{\frac{2k_B T}{m_{\sigma}\omega_{i\sigma}^2}}$ with $m_{Li}$ ($m_{K}$) and $\omega_{iLi}$ ($\omega_{iK}$) being the atom masses and trap frequencies, respectively, and $k_B$ is the Boltzman constant.
Thus, the spin-changing dynamics of $|1\rangle_{Li}+|8\rangle_{K} \rightarrow |6\rangle_{Li}+|1\rangle_{K}$ can be described by the following equations:
\begin{eqnarray}
\frac{d{N}_{|8\rangle_K}}{dt}=-L_{2}\frac{N_{|1\rangle_{Li}}N_{|8\rangle_K}}{V_{LiK}},\label{eq4}
\end{eqnarray}

\begin{eqnarray}
\frac{d{N}_{|1\rangle_K}}{dt}=L_{2}\frac{N_{|1\rangle_{Li}}N_{|8\rangle_K}}{V_{LiK}},\label{eq5}
\end{eqnarray}

\begin{eqnarray}
\frac{d{N}_{|1\rangle_{Li}}}{dt}=-L_{2}\frac{N_{|1\rangle_{Li}}N_{|8\rangle_K}}{V_{LiK}}-L_{2}^{'}\frac{N_{|1\rangle_{Li}}N_{|1\rangle_K}}{V_{LiK}},\label{eq6}
\end{eqnarray}
where $V_{LiK}=\pi^{3/2}\prod_{i=x,y,z} \sqrt{(b_{iLi}^2+b_{iK}^2)}$ is the effective volume. $N_{|1\rangle_K}$, $N_{|8\rangle_K}$ and  $N_{|1\rangle_{Li}}$ are the atom numbers of state $|1\rangle_K$, $|8\rangle_K$ and $|1\rangle_{Li}$, respectively. $L_2$ is the spin-changing rate coefficient of $|1\rangle_{Li}$ and $|8\rangle_K$, while $L_2^{'}$ is proportional to the elastic two-body collision rate of $|1\rangle_{Li}$ and $|1\rangle_K$.

The above equations holds for three reasons: (i) the total atom number of $^{41}$K almost remains a constant during the collision, (ii) lithium atoms that flipped their spin from $|1\rangle_{Li}$ to $|6\rangle_{Li}$ during the collision are ejected from the trap, and (iii) remaining $|1\rangle_{Li}$ atoms gain additional kinetic energy from $|1\rangle_K$ atoms through elastic collision, resulting in further loss of $|1\rangle_{Li}$ atoms. Then, the experimental data of $|1\rangle_K$, $|8\rangle_K$, and $|1\rangle_{Li}$ are fitted using Eqs.~(\ref{eq4})-(\ref{eq6}) simultaneously. Fig.~\ref{figure2} (a)-(b) shows good consistent between the fitting curves (solid lines) and experimental data. The obtained results are $L_2=1.9(2)\times 10^{-12}~\rm{cm^3\cdot s^{-1}}$ and $L_2^{'}=2.4(3)\times 10^{-13}~\rm{cm^3\cdot s^{-1}}$, respectively. We mention that the temperature is considered to be a constant in this fitting, due to small change of temperature from 44.0(6) $\rm{\mu K}$ to 53.4(7) $\rm{\mu K}$ during the HCC.

To compare the experimental measurement with theory, we employ a MQDT calculation to obtain the inelastic scattering length of $^6$Li-$^{41}$K mixture. In this MQDT approach, two quantum defect parameters $K^c_S,~K^c_T$ are required as an input~\cite{Gao1998,Gao2005,Gao2008}. The value of $K^c_S,~K^c_T$ are fine tuned to achieve a best fit on the measured resonance positions in the s-wave channel, and $K^c_{S}=-3.1235,~K^c_{T}=8.9785$ are determined. More details on how to obtain the quantum defects $K^c_S,~K^c_T$ can be found in~\cite{Liu2018}. It's known that the total quantum number m$_F$ of the system should be conserved during the HCC, which determines the corresponding initial and final scattering channels. Figure~\ref{figure3} shows the MQDT predicted scattering length of $|1\rangle_{Li}$-$|8\rangle_K$ channel in the unit of  $\overline{a}$. $\overline{a}=0.956R_{vdw}$ is the mean scattering length~\cite{Chin2010} with $R_{vdw}=40.8a_0$ being the van der Waals length. Since the collision between $|8\rangle_K$ and $|1\rangle_{Li}$ atoms possess elastic and inelastic parts, thus the scattering length can be written as $a_{bf}$=Re($a_{bf}$)$+i$Im($a_{bf}$).  The inelastic scattering length Im($a_{bf}$) accounts for the spin-changing collisions. According to the two-body collision model, the theoretical spin-changing rate coefficient is $L_{2,theo}=2h$|Im($a_{bf}$)|$/\mu _{LiK}$, where $h$ and $\mu_{LiK}$ are Plank's constant and the reduced mass of $^6$Li-$^{41}$K mixture, respectively. By substituting the inelastic scattering length of 0.181(2)$a_0$ (see the inset of Fig.~\ref{figure3}) into this equation, theoretical $L_{2,theo}$ value of $1.45(1)\times10^{-12}~\rm{cm^3\cdot s^{-1}}$ is obtained, which reasonable agrees with the experimental result.

Furthermore, the exact relation between two-body inelastic loss rate and Bose-Fermi contact ($C_{bf}$) can be derived following a similar calculation in ref.~\cite{Braaten2008}:
\begin{eqnarray}
\frac{dN_{|8\rangle_K}(t)}{N_{|8\rangle_K}dt}=-\frac{\hbar |Im(a_{bf})|}{4\pi \mu_{bf} |a_{bf}|^2}C_{|1\rangle_{Li}-|8\rangle_K}(t),\label{eq7}
\end{eqnarray}
where $C_{|1\rangle_{Li}-|8\rangle_K}(t)$ is the contact parameter. We mention that the contact parameter cannot be obtained from the measurement of momentum distribution of the mixture, a common method adopted in previous studies. On the one hand, since the s-wave scattering lengths between the four spin states $|1\rangle_K$, $|8\rangle_K$, $|1\rangle_{Li}$, and $|6\rangle_{Li}$ are comparable, it is difficult to extract the individual $C_{|1\rangle_{Li}-|8\rangle_K}(t)$. On the other hand, it’s extremely difficult to obtain reliable contact parameter in the weakly-interacting regime, simply because the signals are too weak to be observed (the estimated signals are much smaller than the achievable experimental signal-to-noise ratio).  In the contrary, based on Eq.~\ref{eq7}, the measurements of state-to-state HCC allow us to extract the contact parameter of a heteronuclear quantum mixture, a previously inaccessible quantity. Fig.~\ref{figure2} (c) shows the $C_{|1\rangle_{Li}-|8\rangle_K}(t)$ as a function of collision time, which gradually decreases with spin-changing collision. Note that the elastic scattering length is two orders of magnitude larger than the inelastic one, guaranteeing the equilibrium of $|1\rangle_{Li}-|8\rangle_K$ mixture during the HCC. Hence, the obtained $C_{|1\rangle_{Li}-|8\rangle_K}(t)$ can also be regarded as the equilibrium contact of the $|1\rangle_{Li}-|8\rangle_K$ mixture.

For a weakly-interacting system, this result can be further verified using first order perturbation theory and Tan's relation. In the weakly-interacting limit, we have:
\begin{eqnarray}
E\simeq \frac{2\pi \hbar^2 a_{bf}}{\mu_{bf}}\int n_f(\textbf{r})n_b(\textbf{r})d^3r.\label{eq8}
\end{eqnarray}
For a many-body equilibrium state at finite temperature, Tan's adiabatic relation is:
\begin{eqnarray}
\frac{\partial E}{\partial a_{bf}^{-1}}=-\frac{\hbar^2}{8\pi \mu_{bf}}C_{bf}.\label{eq9}
\end{eqnarray}
Therefore, the contact parameter can be determined by the following equation:
\begin{eqnarray}
C_{bf}=16\pi^2a_{bf}^2\int n_f(\textbf{r})n_b(\textbf{r})d^3r.\label{eq10}
\end{eqnarray}
 Using Eq.~\ref{eq10}, the contact parameter is calculated with the input of measured density, which is shown in the Fig.~\ref{figure2} (c). It can be seen that both results agree well with each other, demonstrating the validity of obtaining contact parameter from the measurement of spin-changing dynamics. We emphasis that, the relation that links contact with spin-changing collisions is universal, implying that the method developed here can be applied to other mixtures. For example, in a mixture of alkali-metal and alkaline-earth-metal atoms, the contact parameter can be extracted from the spin-exchange collisions, which is crucial for the study of Kondo effect~\cite{Yao2019a}. One can also apply this method to a system with optical Feshbach resonance~\cite{Chin2010}.

In conclusion, we have intensively studied a specific HCC between $^6$Li and $^{41}$K atoms confined in an optical dipole trap. During this spin-changing collisions, the $^{41}$K atoms flipped their spin from the highest energy state to the lowest one; while for the $^6$Li atoms, the process is just the reverse. The HCC products are directly observed by combining the high-field and mediate-field absorption imaging, providing an excellent opportunity to study the HCC dynamics. The atom number of all the involved states as a function of collision time is further measured, and the spin-changing rate $L_2$ of $1.9(2)\times 10^{-12}~\rm{cm^3\cdot s^{-1}}$ is gained by simultaneously fitting all the two-body loss equations. To verify the experimental results, we have implemented a MQDT calculation to obtain the inelastic scattering length of $|1\rangle_{Li}$-$|8\rangle_K$ scattering channel. The theoretical $L_{2,theo}$ of $1.45(1)\times 10^{-12}~\rm{cm^3\cdot s^{-1}}$ is further calculated, being consistent with the experimental result. Furthermore, the contact parameter of $|1\rangle_{Li}-|8\rangle_K$ mixture is successfully extracted from the measured  spin-changing dynamics with the input of MQDT calculation. The obtained results are particularly important for studying the properties of $^6$Li -$^{41}$K mixture, such as momentum distribution function~\cite{Tan2008a,Braaten2008}, radio-frequency spectroscopy~\cite{Braaten2010}, and dynamic structure factors~\cite{Son2010}. Our work may serve as a foundation for future studies of heteronuclear spin-changing collisions and their dynamics.
\begin{acknowledgements}
We thank Li You, Zhi-Fang Xu, and Youjin Deng for discussions. This work has been supported by the National Key R\&D Program of China (under Grant No. 2018YFA0306501, 2018YFA0306502), NSFC of China (under Grant No. 11874340, 11425417, 11774426), the CAS, the Anhui Initiative in Quantum Information Technologies and the Fundamental Research Funds for the Central Universities (under Grant No. WK2340000081). R. Qi is supported by the Research Funds of Renmin University of China (under Grants No. 15XNLF18, 16XNLQ03).
\end{acknowledgements}

\end{document}